\documentclass[11pt,twoside]{article}
\usepackage{asp2004}
\usepackage{psfig}
\usepackage{epsf}
\usepackage{graphics}
\usepackage{lscape}
\def\edcomment#1{\iffalse\marginpar{\raggedright\sl#1\/}\else\relax\fi}
\reversemarginpar

\begin{document}
\title{DNOs and QPOs in Cataclysmic Variables}
\author{Brian Warner and Patrick A. Woudt}
\affil{Department of Astronomy, University of Cape Town, Rondebosch 7700,
South Africa}

\begin{abstract}
   We describe some of the recent observations of Dwarf Nova Oscillations 
and Quasi-Periodic Oscillations in cataclysmic variable stars. The 
similarities to high frequency QPOs seen in X-Ray binaries make this an 
increasingly useful field to explore. Our principal new result is the 
discovery of 1:2:3 harmonic structure in the DNOs of VW Hyi late in outburst. 
Similar structure is seen in the QPOs of some black hole and neutron star 
X-Ray binaries. In VW Hyi there is a sudden frequency-doubling when the DNO 
period reaches $\sim 39$ s. This first harmonic increases in period to 
$\sim 28$ s at which point the second harmonic appears. Both harmonics 
continue with increasing period, together or separately, with an occasional 
appearance of the fundamental, until the latter reaches $\sim 105$ s, by 
which time VW Hyi has fallen to its quiescent magnitude.
   We also report the first extensive observations of DNOs in a dwarf nova 
(OY Car) in quiescence. These are lpDNOs (longer period DNOs) and show 
similar short time scale properties to the DNOs and lpDNOs seen in 
outburst.
\end{abstract}

\section{Introduction}

The history of rapid oscillations in CVs started exactly 50 years ago with 
the discovery of the 71 s brightness modulation in DQ Herculis by Merle 
Walker in July 1954 (Walker 1956). No further examples were found until 
the advent of digital photometry, and the associated ease of performing 
Fourier transforms (FTs), whereupon very low amplitude oscillations were 
found in many outbursting dwarf novae and some nova-like variables 
(Warner \& Robinson 1972). A recent review gives the total of CVs in which 
such Dwarf Nova Oscillations (DNOs) have been observed as about 50 
(Warner 2004). This enlarged data set confirms most of the earlier general 
correlations, namely

\begin{itemize}
\item{The oscillations are usually only present in high $\dot{M}$ states.}
\item{There is a period-luminosity law, with the minimum of $P_{DNO}$ 
occurring at maximum $\dot{M}$ onto the primary.}
\item{DNOs are maximally coherent at maximum $\dot{M}$, and may 
become incoherent -- in the sense of short trains of coherence 
interrupted by small period changes -- later in outburst.}
\end{itemize}

\section{New Phenomena}

Recent studies have added a number of previously unsuspected 
properties that are making the interpretation of DNOs increasingly complex.
In a series of papers (Woudt \& Warner 2002: WW2; Warner \& Woudt 2002: 
WW1; Warner, Woudt \& Pretorius 2003: WWP; Warner \& Woudt 2004) 
inter alia the following additional behaviours have been noted:

\begin{itemize}
\item{In VW Hyi, in which DNOs are particularly prominent at the very 
ends of outbursts, the usual slow increase of $P_{DNO}$ with falling 
brightness is succeeded by a phase of rapid increase, with $P_{DNO}$ 
doubling in about 5 hours, at the end of which there occurs -- }
\item{Frequency doubling of the DNO, i.e. the appearance of the first 
harmonic, which is in turn superseded by the appearance of the 
second harmonic. During this late phase, which in effect extends 
(at least in the optical) into the first day or so of quiescence, the 
fundamental and 1st and 2nd harmonics can appear in various 
combinations. More details are given below.}
\end{itemize}

\begin{figure}[t]
\plotfiddle{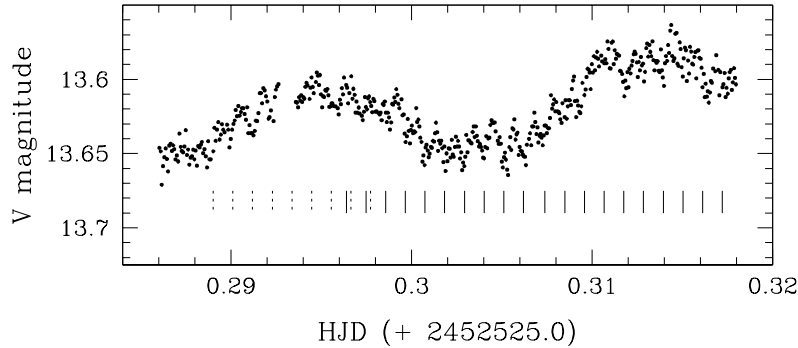}{4.1cm}{0}{70}{70}{-185}{-340}
\caption{The light curve of EC\,2117$-$54.
The lpDNOs with a period of 94.21 s are directly visible in the light curve (from WWP).}
\label{warnerf1}
\end{figure}

\begin{figure}[t]
\plotfiddle{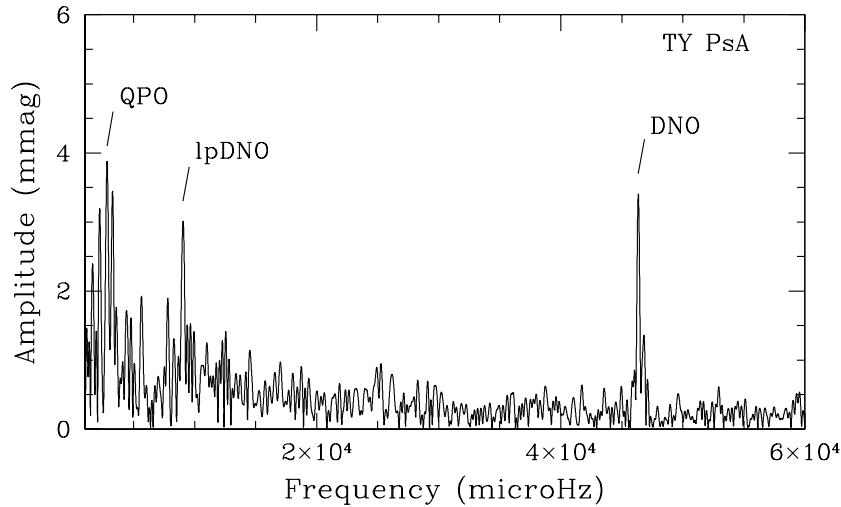}{6.2cm}{0}{62}{62}{-185}{-210}
\caption{The Fourier transform of TY PsA in which the DNO, lpDNO and QPO are present
simultaneously (from Pretorius (2004)).}
\label{warnerf2}
\end{figure}

A completely new type of DNO has been discovered -- the longer period 
DNO (lpDNO) which has a period $\sim 3 - 5$ times that of the DNOs and shows 
little or no dependence on $\dot{M}$. These have been seen in about 17 CVs 
(Warner 2004) but, as with the normal DNOs, are only intermittently 
present. DNOs and lpDNOs behave independently of each other, being seen 
separately or in coexistence and their short time scale variations in period 
or phase appear not to be correlated. Figure~\ref{warnerf1} gives an example of lpDNOs that 
are visible directly in the light curve, and Figure~\ref{warnerf2} shows an example of a FT 
containing both a DNO and a lpDNO.

\begin{figure}[t]
\plotfiddle{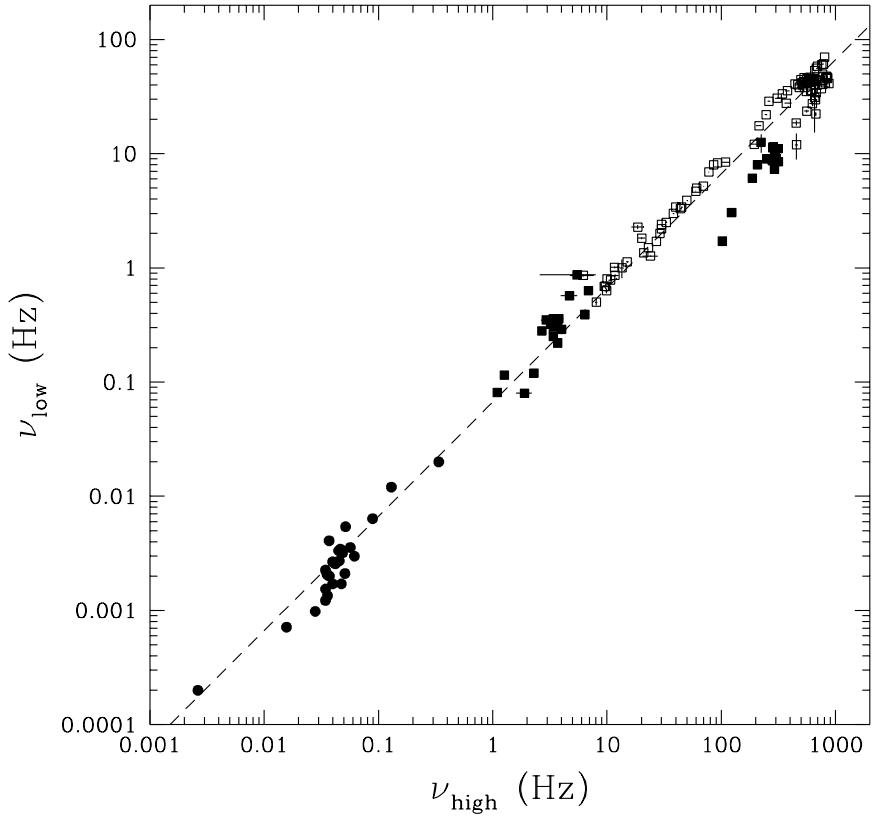}{10.5cm}{0}{62}{62}{-185}{-110}
\caption{The Two-QPO diagram for X-Ray binaries (filled squares: black hole binaries,
open squares: neutron star binaries), and 26 CVs (filled circles). Note that each CV is plotted only
once. The X-Ray binary data are from Belloni, Psaltis \& van der Klis
(2002) and were kindly provided by T.~Belloni.
The dashed line marks $P_{QPO}/P_{DNO} = 15$ (from WWP).}
\label{warnerf3}
\end{figure}

     A third type of rapid modulation in CVs is that called Quasi-Periodic 
Oscillation (QPO), which was first recognised by Patterson, Robinson \& 
Nather (1977), and is of very low coherence, typically changing period or 
phase after only a few cycles. A new appreciation of these oscillations has 
emerged from the realisation that the periods of many QPOs and DNOs are 
related through $P_{QPO} \sim 15 \times P_{DNO}$, as maintained in VW Hyi 
during its rapid deceleration phase (WW2) and seen in numerous other CVs. 
The fact that 
the same relationship is found (Psaltis, Belloni \& van der Klis 1999) in the 
rapid oscillations of X-Ray Binaries (XRBs) shows a connection of 
phenomena across six orders of magnitude in frequency, and links CVs to 
the neutron star and black hole XRBs. The current standing of this 
relationship is shown in Figure~\ref{warnerf3}.

   Occasionally double DNOs are observed, split by the frequency of the 
QPO present at the time.

\section{The Magnetic Model for DNOs and QPOs}

   A magnetic accretion model for DNOs has been developed (Warner 1995; 
WW1), based on an original proposal by Paczynski (1978). In essence it is a 
modification of the standard intermediate polar structure, allowing an 
independently rotating equatorial accretion belt to be set up because the 
magnetic field of the primary is insufficient to couple its outer layers to 
the interior. The optical DNOs are the result of seeing the accretion curtain 
attached to the belt, and/or reprocessing by the disc of the anisotropic 
radiation carried around by the belt. This Low Inertia Magnetic Accretor 
model has had some success in explaining the behaviour of normal DNOs 
(WW1). From this model, however, there is no immediate expectation of 
DNO harmonics, so we need to describe these in detail and suggest how 
further consideration of the LIMA model might explain them.

   The lpDNOs, because of their insensitivity to $\dot{M}$, are hypothesized 
(WW1) to be caused by part of the accretion flow onto the primary being 
channeled along magnetic field lines of the primary itself (which will be 
differentially rotating because of the angular momentum that is fed in at the 
equator, and therefore some variation in period of the lpDNOs is to expected 
according to which field lines are fed from the disc).

   The QPOs are thought to be caused by interruption and reprocessing by a 
traveling wave at the inner edge of the accretion disc, of radiation from the
centre of the system, moving slowly in a 
prograde direction (WW1), for which there is some theoretical expectation 
(Lubow \& Pringle 1993). It is reprocessing of the rotating DNO beam by the 
traveling wave that causes the double DNOs.

\section{Details of the Harmonic Structure in VW Hyi}

   In the earlier analysis of DNO evolution in VW Hyi near the end of 
outburst (WW2) it was pointed out that the rapid increase of $P_{DNO}$, during 
which the period doubles in only five hours, is followed by DNOs that have 
evidently halved in period; but there were insufficient data to analyse this 
fully. In the past year we have concentrated on acquiring more VW Hyi light 
curves during this final phase of outburst, and find that there is a systematic 
effect present that if not fully sampled can lead to the perplexing, apparently 
chaotic variations in period that we had previously deduced.

\begin{figure}[t]
\plotfiddle{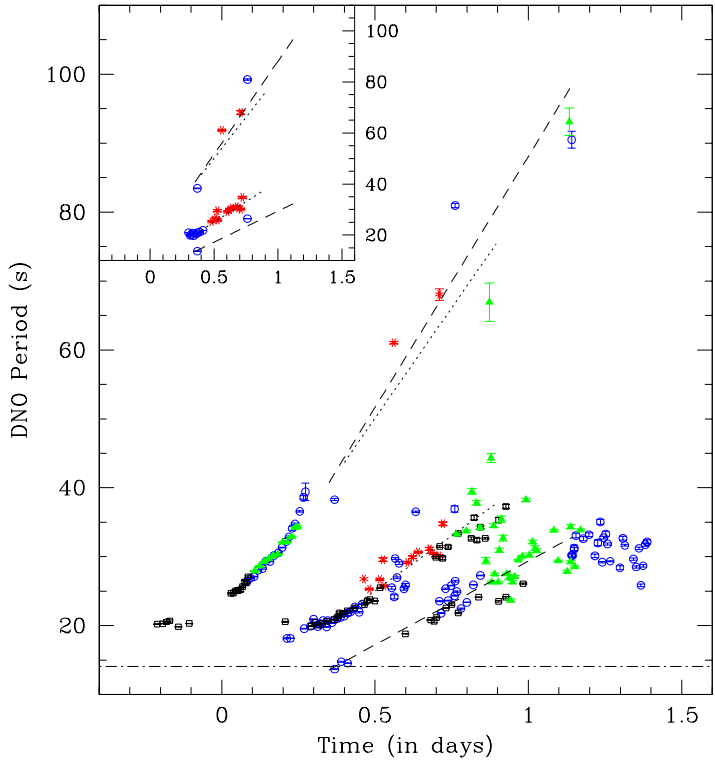}{10.0cm}{0}{55}{55}{-175}{-90}
\caption{The evolution of DNO periods at the end of normal and super outbursts in the dwarf
nova VW Hyi. The different symbols indicate the various different kind of ourbursts (short: asterisk,
normal: open circles, long: open squares, and super outbursts: filled triangles). The dotted and dashed
lines show the result of a least-squares fit to the first and second harmonic, respectively, and are
multiplied by a factor of two and three to show the evolution of the fundamental DNO period. The inset
highlights two observing runs in which the fundamental, first and second harmonic of the DNO period
were present simultaneously. The horizontal dotted-dashed line illustrates the minimum DNO period (14.1 s)
observed at maximum brightness.}
\label{warnerf4}
\end{figure}

   The behaviour of the DNOs is shown in Figure~\ref{warnerf4} (the zero of outburst 
phase is defined in WW2). It should be remembered that what we are 
displaying here are the periodic components derived from Fourier transforms 
(FTs) of subsections of the various observing runs. The light curves were all 
obtained at the Sutherland site of the South African Astronomical 
Observatory, using the UCT CCD Photometer (O'Donoghue 1995) on the 
40-in and 74-in reflectors.

   The steady increase of $P_{DNO}$ from 20 s onwards continues to $\sim 39$ s, 
at which point frequency doubling occurs and what we consider to be the 1st 
harmonic takes over the general increase of period. We have not yet been 
able to observe at the precise moment of transition, but we have bracketed it 
in separate outbursts; it seems to take place in less than $\sim 15$ min and may 
well be almost instantaneous. The 1st harmonic sets out at $\sim 19$ s and 
increases until it reaches $\sim$ 28 s, but the first appearance of an additional 
oscillation, the second harmonic, occurs already when the 1st harmonic is at  
$\sim$ 21 s. 
Initially the 2nd harmonic is rarely seen, but as the periods lengthen it 
eventually becomes the most prominent oscillation, and it persists later in 
the outburst than either of the other components.

\begin{figure}[t]
\plotfiddle{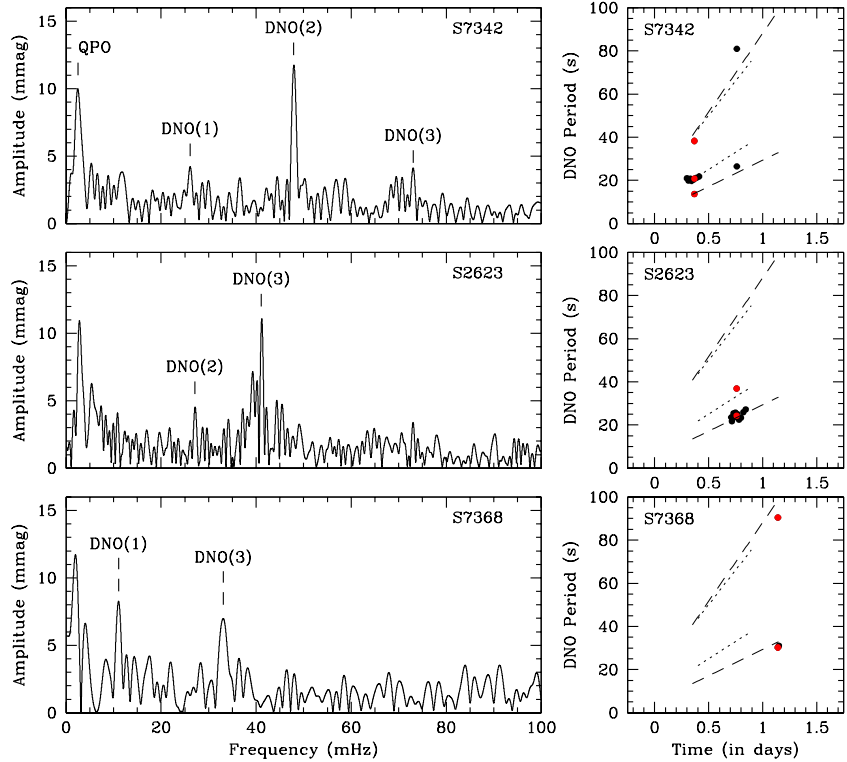}{9.6cm}{0}{62}{62}{-185}{-120}
\caption{Examples of Fourier transforms of VW Hyi in which combinations of the fundamental
and the first two harmonics of the DNO period are present simultaneously.  }

\label{warnerf5}
\end{figure}

       Occasionally both harmonics are present together, and there are very 
occasional appearances of the fundamental. Frequently there are alternations 
between the occurrences of the harmonics. Our FTs show that the 
fundamental and harmonics are in the ratios 1:2:3 within error when they 
appear together. An example of FTs containing the two harmonics is 
shown in Figure~\ref{warnerf5}. When the 1st harmonic has also reached a period of $\sim 40$ s 
it in turn disappears and usually only the 2nd harmonic is present (but with 
infrequent appearances of the fundamental).

\begin{figure}[t]
\plotfiddle{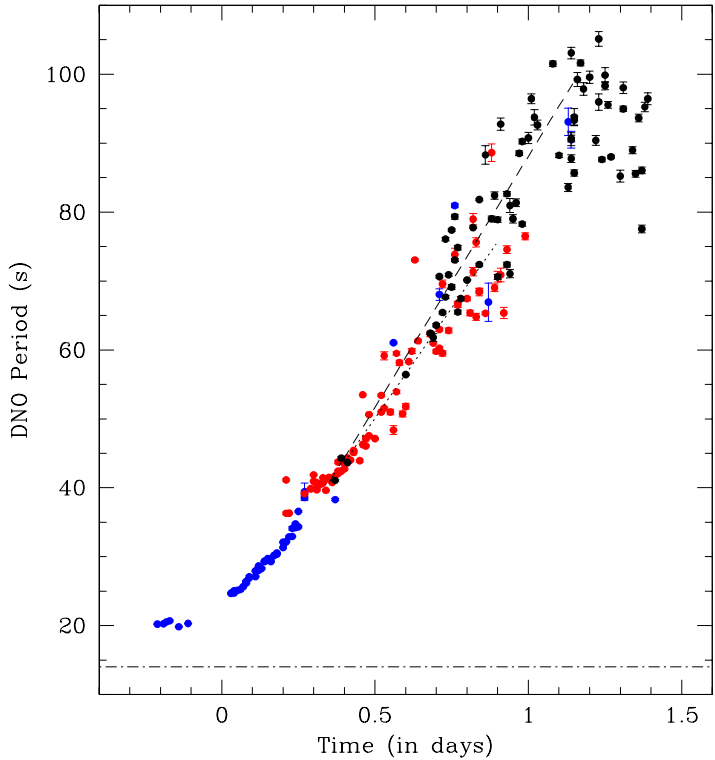}{10.0cm}{0}{55}{55}{-175}{-90}
\caption{The evolution of the fundamental DNO period in VW Hyi at the end of normal and super
outbursts. The periods of the first and second harmonic have been multiplied by a factor of
two and three, respectively. The dotted, dashed and dashed-dotted lines are as in Figure~\ref{warnerf4}.}
\label{warnerf6}
\end{figure}

   Figure~\ref{warnerf6} shows a modified form of Figure~\ref{warnerf4}, where the harmonics have been 
multiplied by appropriate factors of 2 or 3 to generate the implied but 
unobservable fundamental. The fundamental itself is, however, occasionally 
visible, and when it gets to $\sim 80 - 90$ s there is potentially some ambiguity 
with the lpDNOs that have periods near those values in VW Hyi. But the 
lpDNOs are usually pure sine waves, so when any of the harmonics are 
present the period of the fundamental can be predicted precisely and this 
eliminates the confusion. Figure~\ref{warnerf6} shows that the deduced period of the 
fundamental continues to increase steadily, reaching $\sim 105$ s before it 
cannot any longer be detected (we have examined FTs of runs taken shortly after 
that outburst phase without success).

\section{Comparison with XRBs}

    The importance of the 1:2:3 period ratios seen in VW Hyi is that these are 
the first observations in a CV of a phenomenon known for a few years in 
XRBs (McClintock \& Remillard 2004). In particular, the black hole binary 
XTE J1550-564 has strong X-Ray signals at 276 and 184 Hz and a weak 
signal at 92 Hz, which are in the ratio 3:2:1, and GRO J1655-40 has 450 and 
300 Hz oscillations (Remillard et al.~2002). The interpretation of these 
harmonic ratios in XRBs has usually required the tools of General Relativity 
(see review by McClintock \& Remillard (2004)), but it is possible that less 
exotic models might suffice. There certainly should be no recourse to GR to 
explain the harmonic structure in VW Hyi, and we note that the work of 
Robertson \& Leiter (2002, 2003, 2004) shows that magnetically channeled 
accretion, as occurs in CVs, may be possible even from discs around black 
holes.

\section{A Possible Model for the Harmonics in CVs}

  Possible models for the DNO behaviour in VW Hyi will be explored in 
more detail elsewhere, but here we indicate a way in which the harmonics 
may be produced. 

   At maximum of outburst DNOs are rarely seen, but when they are they are 
at 14.1 s, seen both in optical (WW2) and soft X-Rays (van der Woerd, 
Heise \& Bateson 1986). This suggests that the magnetosphere is generally 
squashed down to the surface of the primary, but that occasionally $\dot{M}$ 
reduces sufficiently to allow some magnetically channeled accretion. From 
this, and assuming that only one accreting zone is visible, we can deduce the 
keplerian velocity at the surface of the primary, and hence the mass (0.7  
M$_{\odot}$) of the primary.

Near the end of outburst, $\dot{M}$ onto the primary falls as the disc returns to 
its quiescent state. The ram pressure on the primary's magnetosphere is 
reduced and the inner radius of the truncated disc consequently increases. 
For small inner radii the primary hides the inner parts of disc on its far side; 
but for an inclination of $\sim 63^{\circ}$ and the above white dwarf mass 
the far side becomes visible when the keplerian period of the inner edge 
is $\sim 45$ s. We ascribe the appearance of the first harmonic 
at $P_{DNO} \sim 40$ s to this change in 
geometry -- the ``upper'' accretion curtain is now visible to us on both sides 
of the primary. Possibly also the lower curtain becomes visible at this time, 
accreting $\sim 180^{\circ}$ out of phase with the upper curtain.

   If there is a thickening of the disc in the form of the QPO traveling wall 
then the strongest magnetic field lines sweeping around the inner boundary 
of the disc will accrete at a variable rate, i.e., $\dot{M}$ will be modulated 
at the beat frequency $\omega = \omega_{rot} - \omega_{qpo}$ of the rotation 
frequency $\omega_{rot}$  of the equatorial belt and the frequency 
$\omega_{qpo}$ of the QPO wall. The effect of this modulation on the first 
harmonic of the direct (or disc reprocessed) beam will be to generate frequency 
components $\omega_{rot} +  \omega_{qpo}$, $2 \omega_{rot}$ and 
$3 \omega_{rot} - \omega_{qpo}$. These are close to, but are not exactly in the 
ratios 1:2:3. On the other hand, if what is being observed is DNOs reprocessed 
off the wall itself, then the components will be $\omega_{rot} - \omega_{qpo}$, 
$2 (\omega_{rot} - \omega_{qpo})$ and $3 (\omega_{rot} - \omega_{qpo})$, which have 
exactly the frequency ratios of interest. It is possible that both processes 
could be present simultaneously, where part of the DNO beam is 
reprocessed from the disc and part from the wall. We have some evidence 
that the vertical scatter in Figure~\ref{warnerf4} is caused partly by such an effect. The fact 
that the 2nd harmonic does not make its appearance until the 1st harmonic is 
already in operation is compatible with the hypothesis of rotationally 
modulated $\dot{M}$.

\section{lpDNOs and QPOs in Quiescence }

     From the time of the earliest studies of DNOs it has been realised that 
they rarely if ever occur in quiescence of dwarf novae. However, although 
uncommon, lpDNOs can exist in these low $\dot{M}$ states. The observational 
evidence for quiescent rapid oscillations is discussed in Warner (2004). In 
hard X-Rays, DNOs with periods respectively of $\sim 60$ s, 21.8 s and 33.93 s 
have been seen in VW Hyi (Pandel, Cordova \& Howell 2003: PCH), HT Cas 
(Cordova \& Mason 1984) and SU UMa (Eracleous, Patterson \& Halpern 
1991). The VW Hyi DNOs also appear in soft X-Rays. The HT Cas and SU 
UMa observations are currently the only direct evidence for possible DNOs, 
rather than lpDNOs, in quiescent dwarf novae.

    In the optical, quiescent DNOs with period $\sim 50$ s were reported in OY 
Car and 87.0 s in AQ Eri by WWP. Previously $\sim 100$ s oscillations were 
observed in HT Cas in quiescence (Patterson 1981). In the latter two stars 
the periods are $\sim 4$ times those of their DNOs and therefore fit the notion 
that they are lpDNOs, but our observations of OY Car need more 
explanation.

\begin{figure}[t]
\plotfiddle{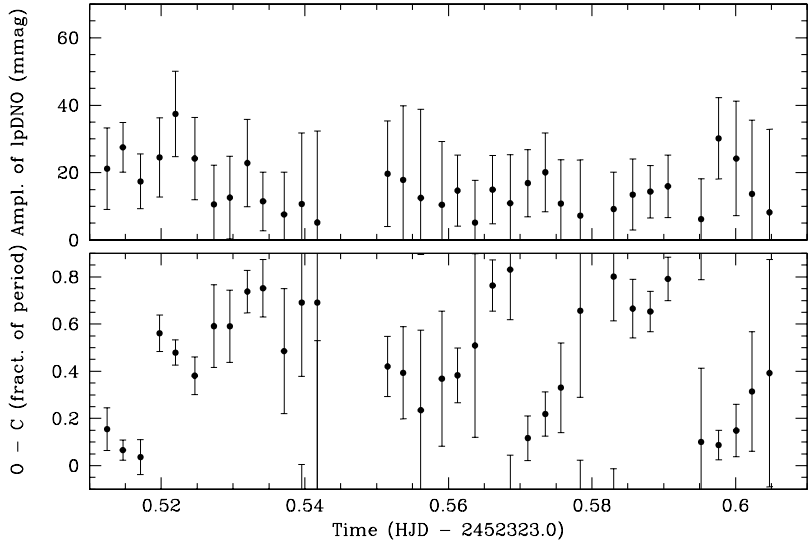}{6.9cm}{0}{61}{61}{-195}{-105}
\caption{An O$-$C diagram of the $\sim$50-s lpDNO harmonic present in OY Car (run S6488).
Note the occasional peaks in amplitude (upper panel) -- around HJD $\sim$ 245\,2323.52,
245\,2323.57 and 245\,2323.60 -- and the relatively high coherence of the lpDNO phase
(lower panel). Each dot represents $\sim 8$ cycles of the 50-s modulation, and there is a 50\%
overlap. There are sudden period increases around 0.520 and 0.560 -- the latter causes a wrap-around
of the phase variation.}
\label{warnerf7}
\end{figure}

    The DNOs in OY Car were illustrated in figure 17 of WWP with 20 mins of light 
curve in which the oscillations were readily visible in the light curve. In fact 
the oscillations were present, but at lower amplitude, throughout the 
remainder of that 2 h run and showed quite large abrupt changes in period 
similar to those seen in outburst. These are illustrated in Figure~\ref{warnerf7}. We find 
that these are the first harmonic of oscillations at $\sim 100$ s -- we see oscillatory 
power also present in the 100 s region of the FT -- and are therefore lpDNOs 
rather than ordinary DNOs. We have found similar lpDNOs in other runs of 
OY Car in quiescence -- they are difficult to detect because of low amplitude 
and the variations in period. We expect that some other quiescent dwarf 
novae, if looked at in the correct way, will also be found to have lpDNOs.

    It is of interest that it is in OY Car that Wheatley \& West (2003) find that 
the hard X-Rays emitted in quiescence are located at a high latitude region 
of the white dwarf surface and from an area small compared with that of the 
white dwarf. This has the signature of magnetic accretion and must be onto 
the body of the white dwarf itself, rather than onto an equatorial belt. 
This is in accord with the hypothesis of WW1.

   The claim by PCH that the X-Ray DNOs in VW Hyi are highly unstable, 
changing period every $\sim$ 2 cycles, is based on the width of the peak in the 
FT. It is more likely that these are lpDNOs, are relatively coherent (as in OY 
Car) and also changed their period during the observation. It has been 
pointed out by Jones \& Watson (1992) and reiterated by Warner (2004) that, 
given the behaviour of DNOs (and lpDNOs) seen in the relatively high 
signal/noise light curves in some optical observations, statistical tests based 
on the assumption of random or stochastic noise are entirely inappropriate; 
in fact, such results are misleading.

\acknowledgments{
BW is supported by research funds from the University of Cape Town; 
PAW is supported by a strategic award to BW from the University, by funds 
from the University and from the National Research Foundation.}

\end{document}